\documentclass[a4paper]{PoS}
\newcommand{\krakow}{Krak\'ow{}}

\title{Control Software for the SST-1M  Small-Size Telescope prototype for the Cherenkov Telescope Array}

\ShortTitle{Control Software for a Small-Size Telescope (SST-1M)}

\author{\firstauthor{V.~Sliusar}{vitalii.sliusar@unige.ch}$^{\,a}$,
R.~Walter\thanks{Speaker.} $^{\,b}$,
C.~Alispach$^{a}$,
I.~Al~Samarai$^{a}$,
W.~Bilnik$^{k}$,
J.~B\l{}ocki$^{c}$,
L.~Bogacz$^{m}$,
T.~Bulik$^{d}$,
F.~Cadoux$^{a}$,
V.~Coco$^{a}$,
D.~della Volpe$^{a}$,
Y.~Favre$^{a}$,
A.~Frankowski$^{g}$,
M.~Grudzi{\'n}ska$^{d}$,
M.~Heller$^{a}$,
M.~Jamrozy$^{e}$,
M.~Janiak$^{g}$,
J.~Kasperek$^{k}$,
K.~Lalik$^{k}$,
E.~Lyard$^{b}$,
E.~Mach$^{c}$,
D.~Mandat$^{l}$,
J.~Micha{\l}owski$^{c}$,
R.~Moderski$^{g}$,
T.~Montaruli$^{a}$,
A.~Neronov$^{b}$,
J.~Niemiec$^{c}$,
T.R.S.~Njoh~Ekoume$^{a}$,
M.~Ostrowski$^{e}$,
P.~Pa{\'s}ko$^{f}$,
M.~Pech$^{l}$,
P.~Rajda$^{k}$,
J.~Rafalski$^{c}$
P.~Schovanek$^{l}$,
K.~Seweryn$^{f}$, 
K.~Skowron$^{c}$,
{\L}.~Stawarz$^{e}$,
M.~Stodulska$^{c}$,
M.~Stodulski$^{c}$,
I.~Troyano Pujadas$^{a}$, 
M.~Wi{\c e}cek$^{k}$,
A.~Zagda\'{n}ski$^{e}$,
K.~Zi{\c e}tara$^{e}$ for the CTA SST-1M Project.\\
\footnotesize{
a.\ DPNC, Universit\'e de Gen\`eve, 24 Quai E.Ansermet, Switzerland.\\
b.\ ISDC, Observatoire de Gen\`eve, Universit\'e de Gen\`eve, Ch. d'Ecogia 16, 1290 Versoix, Switzerland.\\
c.\ Instytut Fizyki J{\c a}drowej im.\ H.\ Niewodnicza{\'n}skiego Polskiej Akademii Nauk, ul.\ Radzikowskiego 152, 31-342 Krak{\'o}w, Poland.\\
d.\ Astronomical Observatory, University of Warsaw, Al.\ Ujazdowskie 4, 00-478 Warsaw, Poland\\
e.\ Astronomical Observatory, Jagiellonian University, ul.\ Orla 171, 30-244, \krakow, Poland.\\
f.\ Centrum Bada{\'n} Kosmicznych Polskiej Akademii Nauk,  18a Bartycka str., 00-716 Warsaw, Poland.\\
g.\ Nicolaus Copernicus Astronomical Center, Polish Academy of Sciences,  Warsaw, Poland.\\
k.\ AGH University of Science and Technology, al.Mickiewicza 30, \krakow, Poland,\\
l.\ Institute of Physics of the Czech Academy of Sciences, 17.\ listopadu 50, Olomouc \& Na Slovance 2, Prague, Czech Republic.\\
m.\ Department of Information Technologies, Jagiellonian University, 
ul.\ prof.\ Stanis{\l}awa {\L}ojasiewicza 11, 30-348  \krakow, Poland.\\
}
}

\abstract{
The SST-1M is a 4-m Davies--Cotton atmospheric Cherenkov telescope optimized to provide gamma-ray sensitivity above a few TeV. 
The SST-1M is proposed as part of the Small-Size Telescope array for the Cherenkov Telescope Array (CTA), the first prototype has already been deployed. The SST-1M control software of all subsystems (active mirror control, drive system, safety system, photo-detection plane, DigiCam, CCD cameras) and the whole telescope itself (master controller) uses the standard software design proposed for all CTA telescopes based on the ALMA Common Software (ACS) developed to control the Atacama Large Millimeter Array (ALMA). Each subsystem is represented by a separate ACS component, which handles the communication to and the operation of the subsystem. Interfacing with the actual hardware is performed via the OPC UA communication protocol, supported either natively by dedicated industrial standard servers (PLCs) or separate service applications developed to wrap lower level protocols (e.g. CAN bus, camera slow control) into OPC UA. Early operations of the telescope without the camera were already carried out. The camera is fully assembled and is capable to perform data acquisition using artificial light source.

}

\FullConference{35th International Cosmic Ray Conference --- ICRC2017\\
		10--20 July, 2017\\
		Bexco, Busan, Korea}

\begin{document}
\section{Introduction}
\label{sec:intro}
The Cherenkov Telescope Array (CTA) will consist of telescopes of three different sizes: Large-Size Telescopes (LSTs), Medium-Size Telescopes (MSTs) and Small-Size Telescope (SSTs). SSTs will be installed only on the southern site of CTA, there will be around 70 of them. The SST-1M is one of three proposed types of Small-Size Telescopes for CTA \cite{ASTRI, GCT} and the only one with a single mirror \cite{SST-1M-design}. The SST-1M is a 4-m Davies--Cotton atmospheric Chereknov telescope optimized to provide gamma-ray sensitivity in a range from 3 to 300~TeV. Due to the big number of these telescopes they will cover a large area, about 7 square kilometers, which is intended to acquire enough statistics of high-energy photons interacting with the atmosphere.

The design of SST-1M is driven by the requirements of CTA for telescope designs: a point spread function for SSTs is 16~mm, pixel angular size $\leq$ 0.25$^{\circ}$ at 4$^{\circ}$ off-axis; the telescope must focus light over 80\% of the required camera field of view diameter with an optical time spread of less than 1.5~ns in standard deviation; a field of view $\geq$ 8$^{\circ}$. The requirements regarding the focal length are also fulfilled due to usage of the Davies--Cotton design, where a 4-m reflector diameter with a focal ratio of 1.4 translates into a focal length of 5.6~m.

One of the key features of SST-1M is the innovative camera. The photo-detection plane (PDP) consists of 1296 pixels, which are custom produced Geiger-mode avalanche photodiodes (G-APDs, or silicon photomultipliers, SiPMs). The pathfinder in this field is the FACT telescope \cite{FACT}, camera of which is composed of G-APDs. The operation of FACT over 6 years revealed the possibility to reliably continue observations even during the half moon nights. The Cherenkov UV light is guided by hollow cones to the G-APDs, preamplified signal of photoelectrons is sent to the DigiCam --- fully digital readout and triggering system \cite{DigiCam}. The use of the DigiCam allows the telescope to reach the dead-time free observations and to have a very robust and flexible subsystem where the behavior of the camera can be easily altered by changing the firmware of the DigiCam rather than changing the hardware elements or camera subsystems.

The control software of the SST-1M complies with officially selected concept of telescope control using ALMA Common Software (ACS) \cite{ACS}. Each subsystem of the telescope is represented by a separate ACS component (the application which is executed internally in ACS), which performs the acquisition of subsystem telemetry data and provides the control over each subsystem. Separate part of the control software is the engineering web-based graphical user interface (web-GUI) which allows an operator to control the telescope in reliable and robust manner, which also provides the overview of all subsystems and their key parameters. Except normal operations when telescope is controlled by high level control software of CTA (array control and data acquisition software or ACTL), it can also perform observations in fully autonomous mode when it is controlled exclusively by the telescope master. The operator in this case can schedule events and telescope master will automatically perform required operation according to predefined procedures while taking care of source tracking, camera control and continuous safety checks.

Early operations of separate subsystems were already carried out using the whole ACS pipeline including the engineering web-GUI, which allows us to perform monitoring, parameters adjustment and full control of the telescope. We will show results on telescope movement and tracking, DigiCam control and data readout, photo-detection plane telemetry readout and control, mirrors alignments, image readout from CCD cameras, safety system tests.

\begin{figure}[b]
	\centering
	\includegraphics[width=0.8\textwidth]{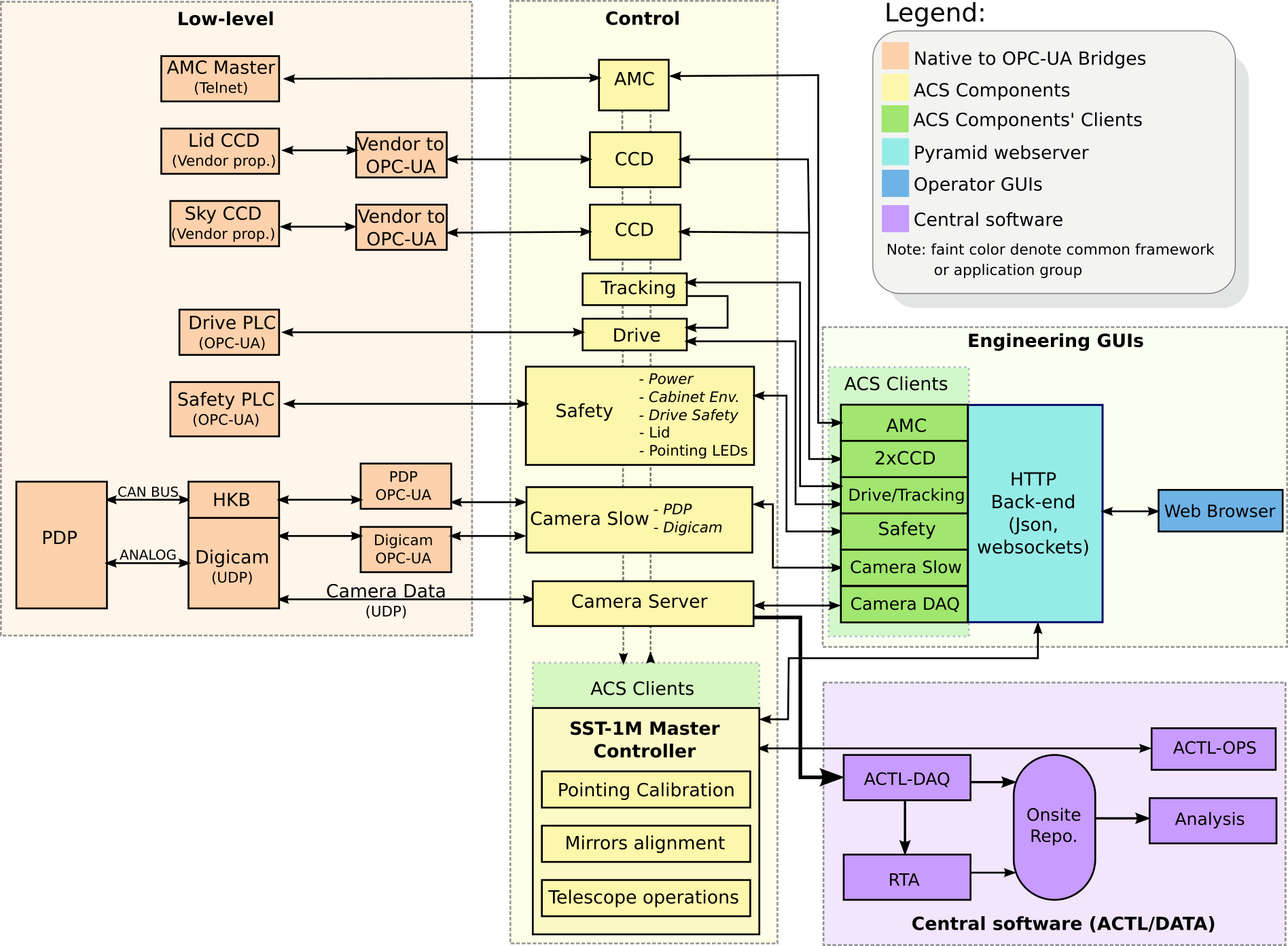}
	\caption{Logic diagram of the software subsystem communication for SST-1M using ACS-centric approach.}
	\label{fig:logic}
\end{figure}

The list and description of all telescope subsystems is given in Section~\ref{sec:subsystems} with a logic diagram of communication between different software and hardware elements (Figure~\ref{fig:logic}). The telescope master is described in Section~\ref{sec:master}. The web-GUI and its use is provided in Section~\ref{sec:gui}.

\section{Telescope subsystems}
\label{sec:subsystems}

The SST-1M includes the following hardware: active mirror control (AMC), which includes the AMC wireless ZigBee controller and two actuators on each of 18 mirrors; a lid CCD camera for pointing, bending model calibration and automatic mirrors adjustments; a sky CCD camera for pointing calibrations; telescope drives, a drive PLC, and drive controllers, which perform slewing and tracking of gamma-ray sources; PDP, a separate subsystem of the camera which detects the Cherenkov UV light and preamplifies the signal for further processing; DigiCam, which performs the readout of preamplified signal, digitalizes it, triggers the data acquisition in case of extensive-air-showers events and sends the data to the camera server; a safety PLC, which continuously monitors the environment conditions and has the internal safety procedures in case of any failures or malfunctions. Communication with the DigiCam and PDP is performed via the UDP by separate dedicated software which then provides the higher-level access for ACS via OPC UA, a standardized industrial protocol for system automation. The same approach is used for the CCDs where the vendor specific API is used inside a dedicated application which later provides an OPC UA interface eventually used by ACS and ACS components of each subsystem.
The control software of the SST-1M is written mostly in Java, Python and C/C++. In Java: ACS components of all subsystems (except telescope master and data acquisition ACS component) and OPC UA bridges for CCD cameras, DigiCam. In Python: ACS components clients for engineering web-GUI and telescope master, telescope master itself and PDP OPC UA bridge. C/C++ is used for data acquisition software only. 

\textbf{CCD cameras}. Both lid and sky CCD cameras use the OPC UA to Allied Vision vendor API bridges provided by MST telescope team. Respective ACS components are connected via OPC UA and allow ACS components clients to change the gain, exposure time, acquisition frames rate, color depth, etc. and to start the data acquisition.

\textbf{AMC}. The active mirror control consists of wireless communication master which provides the telnet access with AT-like commands interface, and proxies the request from ACS via the ZigBee wireless protocol to the actuators' controllers which then command actuators to perform an appropriate action or send the requested data.

\textbf{Photo-detection plane}. Each of three independent sectors of PDP is connected via the CAN to one of three trigger boards of the DigiCam, firmware of the DigiCam acts as a UDP-to-CAN proxy and provides a transparent bridge over the TCP/IP. The OPC UA bridge has an implementation of a CAN interface which encodes the high-level requests and sends them via the UDP to the DigiCam and the answer from PDP is obtained.

\textbf{DigiCam}. The DigiCam uses a specially designed protocol for control and monitoring, this protocol is implemented inside the OPC UA bridge. The bridge then exposes all the data points for each telemetry parameter of the DigiCam to the DigiCam ACS component via the OPC UA.

\textbf{Safety PLC}. The safety PLC provides native OPC UA server, which is directly used by the safety PLC ACS component. This ACS component is based on asynchronous approach and informs the telescope master about any change in any data point.

\textbf{Drive PLC}. The drive PLC provides native OPC UA server, which is directly used by the drive PLC ACS component. This ACS component was provided by the MST team in terms of the effort of software unification among different telescope teams inside CTA.

\section{Telescope master}
\label{sec:master}
The telescope master is an ACS component written in Python. It provides the autonomous and ACTL controlled mode of telescope operations. The telescope master is using specially written Python modules to control and acquire data from different subsystems. These modules are ACS component clients and are also used in the web-GUI. They use asynchronous callback approach to inform master about any changes in the data points in each subsystem, which leads to instant reaction of the master to any kind of event or error. All these modules use a generic ACSWorker and ACSDatapoint classes. These classes allow software to connect to and control any ACS component including the possibility of asynchronous interaction and notifications also inside the module itself.

The SST-1M telescope has three control modes: \textit{Local}, when telescope is operated manually by local personnel of the CTA site; \textit{GUI}, when the maintenance is performed remotely or locally using the engineering web-GUI; \textit{Remote}, when the telescope is either controlled by telescope master directly or while controlled by by high level control software of CTA.

\subsection{Telescope state machine}
\begin{figure}[t]
	\centering
	\includegraphics[width=0.7\textwidth]{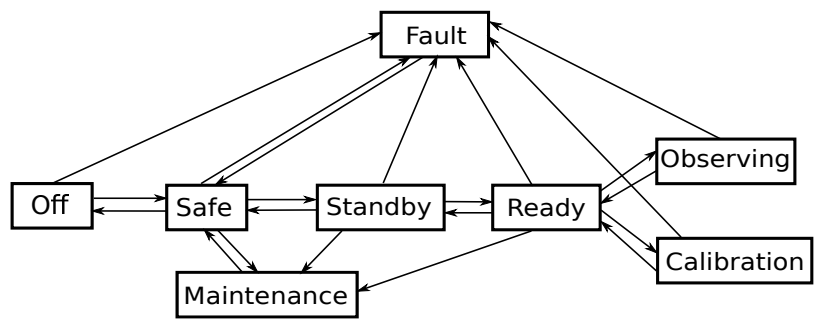}
	\caption{Telescope state machine during normal (non-maintenance) operations either in autonomous or in ACTL controlled mode. }
	\label{fig:statemachine}
\end{figure}

During normal operations the SST-1M telescope exists in one of multiple allowed states according to the telescope state machine (Figure~\ref{fig:statemachine}). The data acquisition during the observations of gamma-ray sources is performed either in the \textit{Science} or in the \textit{Calibration} states. While idle or during daytime the telescope is in the \textit{Safe} state. The telescope state is changed either based on the schedule or performing the CTA high level control software request, the \textit{Fault} state can be reached from any other state when critical error or malfunction occurs, the \textit{Maintenance} state can only be reached manually by changing the telescope control mode from \textit{Remote} to \textit{GUI}.

\begin{itemize}
\item \textit{Off}: In this state the telescope is considered powered off, all subsystems except the safety PLC are powered off and the telescope is locked. The safety PLC is always running to ensure telescope safety and allow operator to turn on all subsystems. The telescope master is only connected to the safety PLC ACS component to monitor it and in case of necessity to change the state to the \textit{Safe}. 
\item \textit{Safe}: The basic state of the telescope and telescope master. All subsystems except the safety PLC are turned off, so no connection to any other ACS component is possible. The telescope is automatically parked if this state is reached from the \textit{Standby} state.
\item \textit{Standby}: The safety PLC is in the \textit{Standby} state, the telescope is locked, the drive PLC is activated. The telescope master establishes connection to the drive PLC.
\item \textit{Ready}: The telescope master acquires establishes the connection to the DigiCam slow control ACS component and to the PDP ACS components. Connections to lid and sky CCDs ACS components are established and the telescope starts to get the images from both CCDs. 
\item \textit{Observing}: The drive system is initialized, the telescope is unlocked and is ready to slew, tracking is allowed. The PDP common high voltage and high voltage for pixels are now turned on so the telescope now starts to convert UV photons from extensive air showers into electric signal. The DigiCam is commanded to start triggering the data acquisition. The camera's lid is open.
\item \textit{Calibration}: Depending on calibration type the connection can be established to any the telescope subsystems.
\item \textit{Maintenance}: This is the manually initialized state, this state cannot be reached in \textit{Remote} control mode of the telescope, either \textit{Local} or \textit{GUI} mode should be activated first.
\item \textit{Fault}: The telescope can reach this state automatically based on the safety PLC procedures or by telescope due to master error report to the safety PLC. If state was reached automatically by the safety PLC, the telescope master requests the telescope log which becomes then available to ACTL.
\end{itemize}

\begin{figure}[t]
	\centering
	\includegraphics[width=0.9\textwidth]{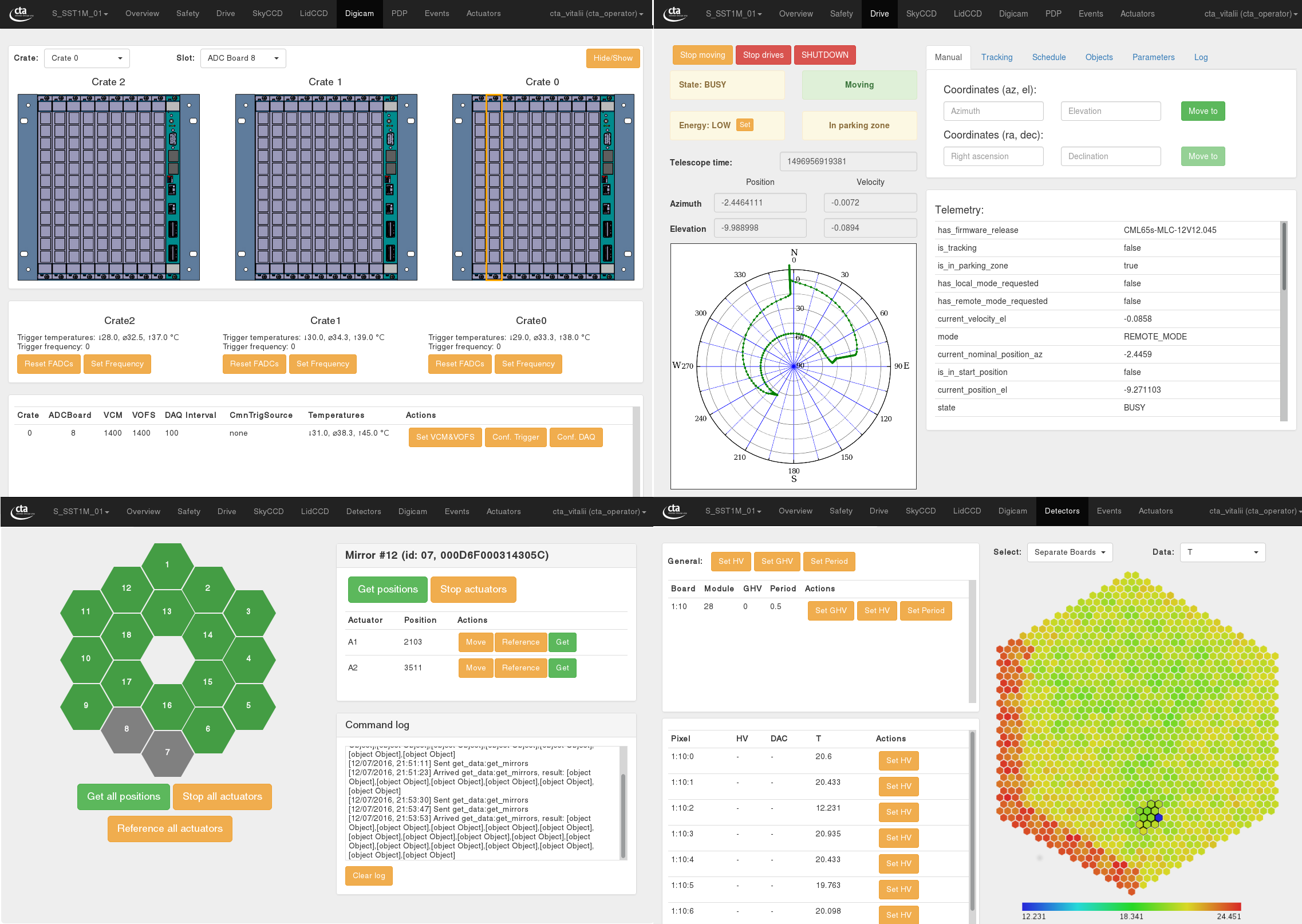}
	\caption{Web-based engineering graphical user interface for the DigiCam (top left), drive system (top right), active mirror control system (bottom left) and PDP (bottom right). }
	\label{fig:gui}
\end{figure}

\section{Engineering web-GUI }
\label{sec:gui}
The engineering graphical user interface allows an operator to control and monitor each subsystem of the telescope via usual web browser. The frontend part of the web-GUI is using jQuery, HTML5, Twitter Bootstrap and socket.io to process and display data obtained from the backend web-GUI server. The backend is using the same Python ACS component clients which are used by telescope master. This clients provide a generic interface to any ACS components and allow transparently transfer data and metadata (units, resolution, description) of each data point. As a web-server the multithreaded Python application based on Pyramid framework is created. The data between the frontend and backend is sent via websockets, the web page is subscribing to the predefined list of data points and obtains the values from ACS as soon as ACS internally is updating the data from respective OPC UA server of each subsystem. The data points are pushed by the ACS component clients via websocket to the web page, this approach allows the data to be send only when it is actually updated which saves the bandwidth and significantly decreases the lag between value update inside the subsystem and on the web-GUI. The system allows multiple authentications in either of two modes: read-only, when the web page is only capable of displaying data without any possibility to execute any methods, i.e. control the telescope; read-write mode, which is used by operator to operate the telescope either for observations or for maintenance purposes.

\section{Telescope tests}
\label{sec:tests}

Using the software, described in the current paper, we were able to perform the following successful tests:
\begin{itemize}
\item Telescope slewing tests in allowed ranges: from $-270$ to $270$ degrees in azimuth and from 0 to 91 degrees in elevation.
\item Tracking of the multiple stars with simultaneous images being taken by both CCD cameras.
\item Continuous readout of all parameters and control of the PDP.
\item Continuous readout of all parameters and control of the DigiCam.
\item The safety PLC telescope state changes, power control of separate subsystems and continuous readout from all sensors.
\item Checks of safety procedures and verification of automatic error responses of each subsystem of the SST-1M.
\item Data acquisition using the PDP and the DigiCam induced by artificial light source in the lab.
\item Mirrors' alignment \cite{bokeh}.
\end{itemize}
Long tests (about a week long) were performed with the safety PLC, when the telescope master and engineering graphical user interface were continuously obtaining and displaying data from the safety PLC. The safety PLC in Krakow was successfully connected using RS-485 interface bridged over Internet to housekeeping board in Geneva. This board, while the telescope is in any state except the \textit{Off} state, is running and provides continuous readout of humidity, air pressure and temperature in multiple points inside the camera, this data is used by the safety PLC to ensure safety of the camera according to the predefined thresholds and safety procedures.

\section{Conclusions}
Having innovative features, the SST-1M telescope proposed for the Cherenkov Telescope Array is foreseen to be a high performance instrument for high-energy astrophysics observations due to dead-time free robust camera and easily scalable control software, which allows an operator to control telescope in fully autonomous mode or CTA driven mode. The generic approach used in this software also allows it to be reused for other telescopes of CTA and to create slow control logging systems and alarm systems. Control software includes innovative web-GUI which can be used by an operator to control a telescope either for observations or for maintenance purposes.

\section{Acknowledgments}
We gratefully acknowledge support from the agencies and organizations listed under Funding Sources at this website: \textit{http://www.cta-observatory.org}. In particular we are grateful for support from the NCN grant DEC-2011/01/M/ST9/01891 and the MNiSW grant 498/1/FNiTP/FNiTP/2010 in Poland. This work was conducted in the context of the SST-1M CTA Project.

\nocite{*}
\bibliographystyle{aipnum-cp}%
\bibliography{icrc_sliusar_v0.1}%

\begin{thebibliography}{99}
\bibitem{FACT} 
H.~Anderhub \emph{et al.}, \emph{Design and Operation of FACT - The First G-APD Cherenkov Telescope}, JINST \textbf{8} (2013) [arXiv:1304.1710].



\bibitem{DigiCam}
P.~Rajda \emph{et al.}, \emph{DigiCam - Fully Digital Read-out and Trigger Electronics for the SST-1M Telescope proposed for the Cherenkov Telescope Array}, in proceedings of \emph{34th International Cosmic Ray Conference (ICRC2015)} (2015) [arXiv:1508.05894].

\bibitem{ASTRI}
R. Canestrari for the ASTRI and CTA Consortium, \emph{The ASTRI SST-2M prototype for the next generation of Cherenkov telescopes: structure and mirrors}, in proceedings of \emph{SPIE} \textbf{8861} (2013) id 886102 13 pp.

\bibitem{GCT}
L.~Tibaldo for the CTA Consortium, \emph{The gamma-ray Cherenkov telescope for the Cherenkov telescope array}, in proceedings of \emph{AIP Conference} \textbf{1792} 080004 (2017).

\bibitem{SST-1M-design}
J.~Niemiec \emph{et al.} for the SST-1M sub-Consortium (SST-1M sub-Consortium), \emph{Prototype of the SST-1M Telescope Structure for the Cherenkov Telescope Array}, in proceedings of \emph{34th International Cosmic Ray Conference (ICRC2015)} (2015) [arXiv:1509.01824].

\bibitem{ACS}
G.~Chiozzi, \emph{et al}, \emph{CORBA-based Common Software for the ALMA project}, Advanced Telescope and Instrumentation Control Software II. Edited by Lewis, Hilton. Proceedings of the \emph{SPIE} \textbf{4848} (2002) pp. 43-54.

\bibitem{bokeh}
M.~Ahnen, \emph{et al}, \emph{Bokeh mirror alignment for Cherenkov telescopes}, Astroparticle Physics \textbf{82} (2016) pp. 1-9.



\end{thebibliography}

\end{document}